\RequirePackage{lineno}
\documentclass[prd,preprint,tightenlines,floatfix,
showpacs,preprintnumbers,nofootinbib,eqsecnum,superscriptaddress]{revtex4}

\usepackage[T1]{fontenc}		% fnot encoding with polish letters !or OT4
\usepackage{kpfonts}
\usepackage{amsmath,amsfonts,amssymb,amstext,mathrsfs}
\usepackage{mathpazo}

\usepackage[dvips]{graphicx}
\usepackage{epsf,float}
\usepackage{revsymb}

\usepackage{soul}

\usepackage{dcolumn}% Align table columns on decimal point
\usepackage{braket}
\usepackage{color,xcolor}
\usepackage{graphicx}
\usepackage{subfigure}
\usepackage{multirow}
\usepackage{tabularx}
\usepackage{pstricks}
\usepackage[section]{placeins}
\usepackage{booktabs}
\usepackage{array}
\usepackage{blindtext}

\usepackage{hyperref}
\begin{document}

%\linenumbers
%\Blindtext

\title{Light-by-light scattering in ultra-peripheral heavy-ion collisions \\ at low diphoton masses}

\author{Mariola K{\l}usek-Gawenda}
 \email{Mariola.Klusek@ifj.edu.pl}
\affiliation{Institute of Nuclear Physics Polish Academy of Sciences, Radzikowskiego 152, PL-31-342 Krak\'ow, Poland}

\author{Ronan McNulty}
 \email{Ronan.Mcnulty@cern.ch}
\affiliation{School of Physics, University College Dublin, Dublin, Ireland}

\author{Rainer Schicker}
 \email{schicker@physi.uni-heidelberg.de}
\affiliation{Physikalisches Institut, Ruprecht-Karls-Universit{\"a}t Heidelberg, Heidelberg, Germany}

\author{Antoni Szczurek
\footnote{Also at \textit{Faculty of Mathematics and Natural Sciences, University of Rzesz\'ow, Pigonia 1, PL-35-310 Rzesz\'ow, Poland}.}}
\email{Antoni.Szczurek@ifj.edu.pl}
\affiliation{Institute of Nuclear Physics Polish Academy of Sciences, Radzikowskiego 152, PL-31-342 Krak\'ow, Poland}

\today

\begin{abstract}
%----------------------------

We present a study of photon-photon scattering in the mass range
$W_{\gamma\gamma} < 5$ GeV.
We extend earlier calculations of
this cross section for $W_{\gamma\gamma}>$ 5 GeV into the low mass range where
photoproduction of the pseudoscalar resonances $\eta$, $\eta^{'}(958)$
contributes to two-photon final states. We present the elementary photon-photon
cross section as a function of diphoton mass $M_{\gamma\gamma}$ arising from lepton and quark
loop diagrams, and the visible cross section 
obtained with the gamma-gamma decay
branching fractions of the resonances $\eta$, $\eta^{'}(958)$, $\eta_c(1S)$, $\eta_c(2S)$,
$\chi_{c0}(1P)$. We derive the corresponding cross sections
in ultra-peripheral Pb-Pb collisions at $\sqrt{s_{NN}}$ = 5.02 TeV by folding the
elementary cross section with the heavy-ion photon fluxes. We consider the
dominating background of the two photon final state  which arises from gamma
decays of photoproduced $\pi^{0}$-pairs. Such $\pi^{0}$-pairs contribute to the
background when only two of the four decay photons are within the experimental
acceptance, while the other two photons escape undetected. 
We reduce this background by applying cuts on 
	asymmetries of transverse momenta of the two photons 
and indicate how the background can be further suppressed using a multivariate sideband analysis.	
We present the cross section for the signal and the
background at midrapidity $|\eta| <$ 0.9, 
and in the forward rapidity range
2.0 $< \eta <$ 4.5.   

\end{abstract}

%\pacs{25.75.-q,25.75.Dw,13.60.Rj,13.90.+i}

\maketitle

%----------------------------
\section{Introduction}
\label{sec:intro}
%----------------------------

The properties of light have at all times fascinated physicists. The
development of instruments for measuring light, and the design of
experiments using light, have resulted in fundamental contributions to our
current knowledge of modern physics. 
%\textbf{As examples, we mention here the development of the telescope by Galileo Galilei for observing celestial objects, and the Michelson-Morley experiment for testing  the aether hypothesis. 	The study  of planetary orbits resulted in the formulation of Keplers laws of planetary motion, and the outcome of the Michelson-Morley experiment negated the aether hypothesis, thereby establishing the foundation for formulating special relativity. }
Our present understanding of the behaviour of light
at the classical level is conveniently expressed by Maxwell's equations.
These equations are linear in the electric currents as sources, and in
the resulting electric and magnetic fields {\bf E} and {\bf B}. At the classical level, two
electromagnetic waves in vacuum will superimpose, and will pass through
each other without scattering. With the emergence of quantum mechanics,
first attempts to formulate equations for the scattering of photons off
each other were formulated in 1925~\cite{Schaposchnikow:1925}. Shortly
thereafter, Louis de Broglie associated possible solutions of these
equations to non-trivial scattering between two photons, in violation of
the superposition principle \cite{Broglie:1926}. This breakdown of the
superposition principle can be incorporated in non-linear Lagrangians,
which result in non-linear field equations. Attempts to solve
the infinite Coulomb energy of a point source led to Born-Infeld
electrodynamics, which effectively confirms the linearity down to some
length scale, and introduces non-linearities at smaller
scales \cite{Born:1925}. Virtual electron-positron pairs were suggested
in 1933 to be at the origin of this photon-photon
scattering \cite{Halpern:1933}. The continued study of photon-photon
scattering led to the Euler-Kockel-Heisenberg Lagrangian which modifies
the classical Maxwell's equation in vacuum by leading non-linear
terms \cite{Heisenberg:1936}. A more detailed account of the history of
photon-photon scattering can be found in a recent
review \cite{Scharnhorst:2017wzh}.

The advent of accelerating heavy-ions to ultra-relativistic energies
with large associated photon fluxes provides new opportunities in gaining
insight into the mechanisms of photon-photon
scattering \cite{dEnterria:2013zqi,Klusek-Gawenda:2016euz}.
At these energies, not only lepton and quark loop diagrams constitute the
signal, but also meson-exchange currents are predicted to
contribute \cite{Klusek-Gawenda:2013rtu}. Such measurements in heavy-ion
collisions hence extend photon-photon scattering from a pure QED issue
to a subject connecting QED with QCD topics.
This QED-QCD connection is, for example, an important ingredient 
in the quantum electrodynamical calculation of the  muon anomalous moment
(see e.g. \cite{Bennett:2006fi}, \cite{Colangelo:2017urn}  and references
therein). In these calculations, light-by-light diagrams appear inside more
complicated QCD-QED diagrams. There are proposals to study such diagrams  in
double back-Compton scattering using high-power lasers \cite{Bula:1996st}.

First evidence of diphoton measurements in ultra-peripheral heavy-ion
collisions have been reported by the ATLAS and CMS
Collaborations \cite{Aaboud:2017bwk,Sirunyan:2018fhl}. These data are,
however, restricted to photon-photon invariant masses W$_{\gamma\gamma} >$
5 and 6 GeV for the CMS and ATLAS analysis, respectively.
ATLAS measured a fiducial cross section of $\sigma = 70 \pm 24$ (stat.)
$\pm 17$ (syst.) nb and theoretical calculations
(including experimental acceptance) gave  $45 \pm 9$ nb
\cite{dEnterria:2013zqi} and $49 \pm 10$ nb \cite{Klusek-Gawenda:2016euz}. 
ATLAS comparison of its experimental results to the predictions
from Ref.~\cite{Klusek-Gawenda:2016euz} shows a reasonable agreement. 
Only 13 events were identified in the ATLAS data sample, 
with an expectation
of 7.3 signal events and 2.6 background events from Monte Carlo
simulations \cite{Aaboud:2017bwk}.
Recently, the CMS Collaboration measured the same process but for a slightly lower threshold of diphoton invariant mass \cite{Sirunyan:2018fhl}.
The measured fiducial light-by-light scattering cross section, $\sigma = 120 \pm 46$ (stat.) $\pm 28$ (syst.) $\pm 4$ (theo.) nb 
was obtained.
The CMS measured value is in good agreement with the result 
obtained according to 
Ref.~\cite{Klusek-Gawenda:2018ijg}. It is important to further test the light-by-light scattering - for different energies and with a better precision.

The purpose of the study presented here is to examine the possibility  of
measuring photon-photon scattering in ultra-peripheral heavy-ion collisions
at LHC energies in the range $W_{\gamma\gamma}< 5$~GeV.
At lower diphoton masses, 
photoproduction of meson resonances plays a role
in addition to the Standard Model box diagrams \cite{Lebiedowicz:2017cuq},
as well as double photon fluctuations into light vector mesons \cite{Klusek-Gawenda:2016euz} 
or two-gluon
exchanges \cite{Klusek-Gawenda:2016nuo}.

In the present study we consider also background 
from the $\gamma \gamma \to \pi^0 (\to \gamma \gamma) \pi^0 (\to \gamma \gamma)$ process measured e.g. by the Belle \cite{Uehara:2009cka} 
and Crystal Ball \cite{Marsiske:1990hx} collaborations. 
In Ref.~\cite{Klusek-Gawenda:2013rtu} a multi-component model,
which describes 
the Belle $\gamma \gamma \to \pi^0 \pi^0$ data,
has been constructed.
This model was used next to make predictions for the
$A A \to A A \pi^0 \pi^0$ reaction \cite{Klusek-Gawenda:2013rtu}. 
If only two photons from different neutral pions are measured 
within the experimental acceptance
such an event could be wrongly identified as $\gamma \gamma \to \gamma \gamma$ scattering.
Extra cuts need to be imposed to reduce or eliminate
	this background.

This paper is organized as follows. 
In section 2, the theoretical framework for calculating photon-photon scattering is outlined, the background from $\gamma \gamma \to \pi^0\pi^0$ is reviewed, and the cross section
	for photoproduction of resonances is studied. 
 The experimental acceptances used for the given
photon-photon scattering cross sections are described in section 3.
The nuclear cross sections for light-by-light scattering in ultra-peripheral lead-lead collisions at the energy $\sqrt{s_{NN}}=5.02$~TeV
are given in section 4.
The treatment of the $\pi^{0}\pi^{0}$ background is discussed in section 5,
and conclusions are presented in section 6.

%******************************

%For LHCb one can also expect a background from
%the $Pb Pb \to Pb Pb e^+ e^-$ where both $e^+$ and $e^-$ are
%misidentified as photons.

%\textbf{Ronan's comment: Delete last sentence of first paragraph.\\
%	delete last sentence in section 1.}

%----------------------------------------------------------------------------
\section{Theoretical approach for the signal and background}
\label{sec:theo}
\begin{figure}[!h]
	(a)\includegraphics[scale=0.35]{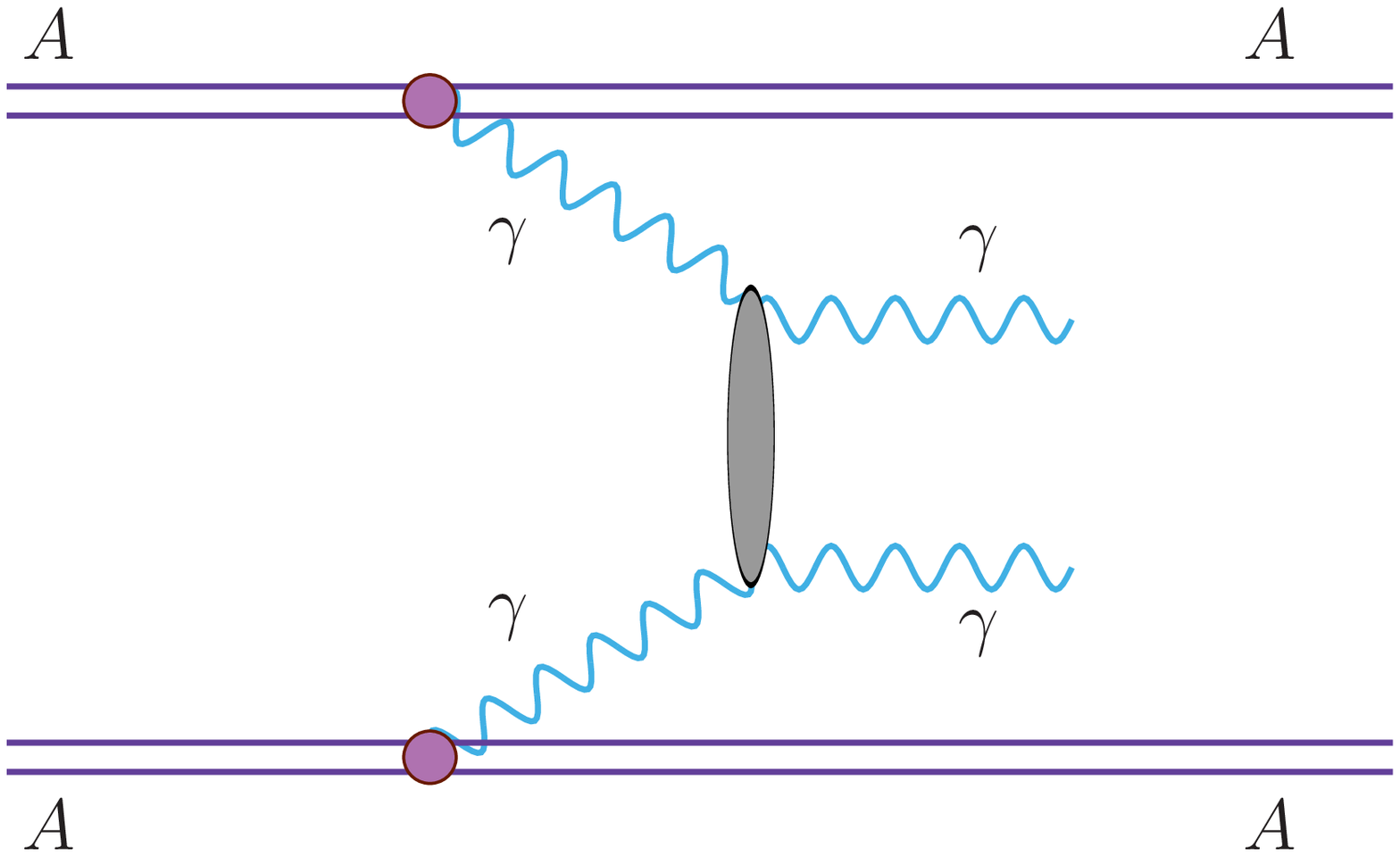} \hspace*{1.cm}
	(b)\includegraphics[scale=0.35]{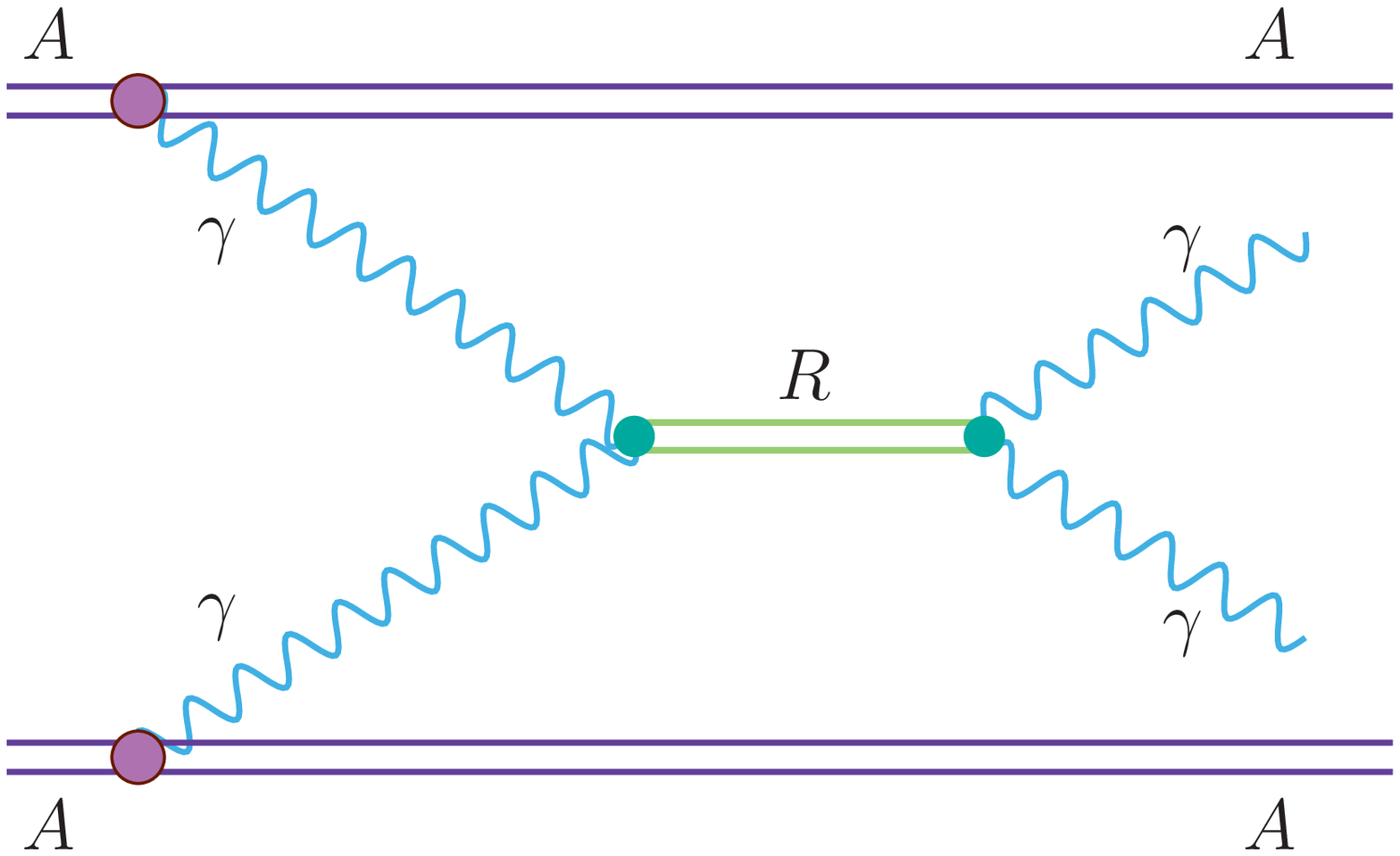}
	(c)\includegraphics[scale=0.35]{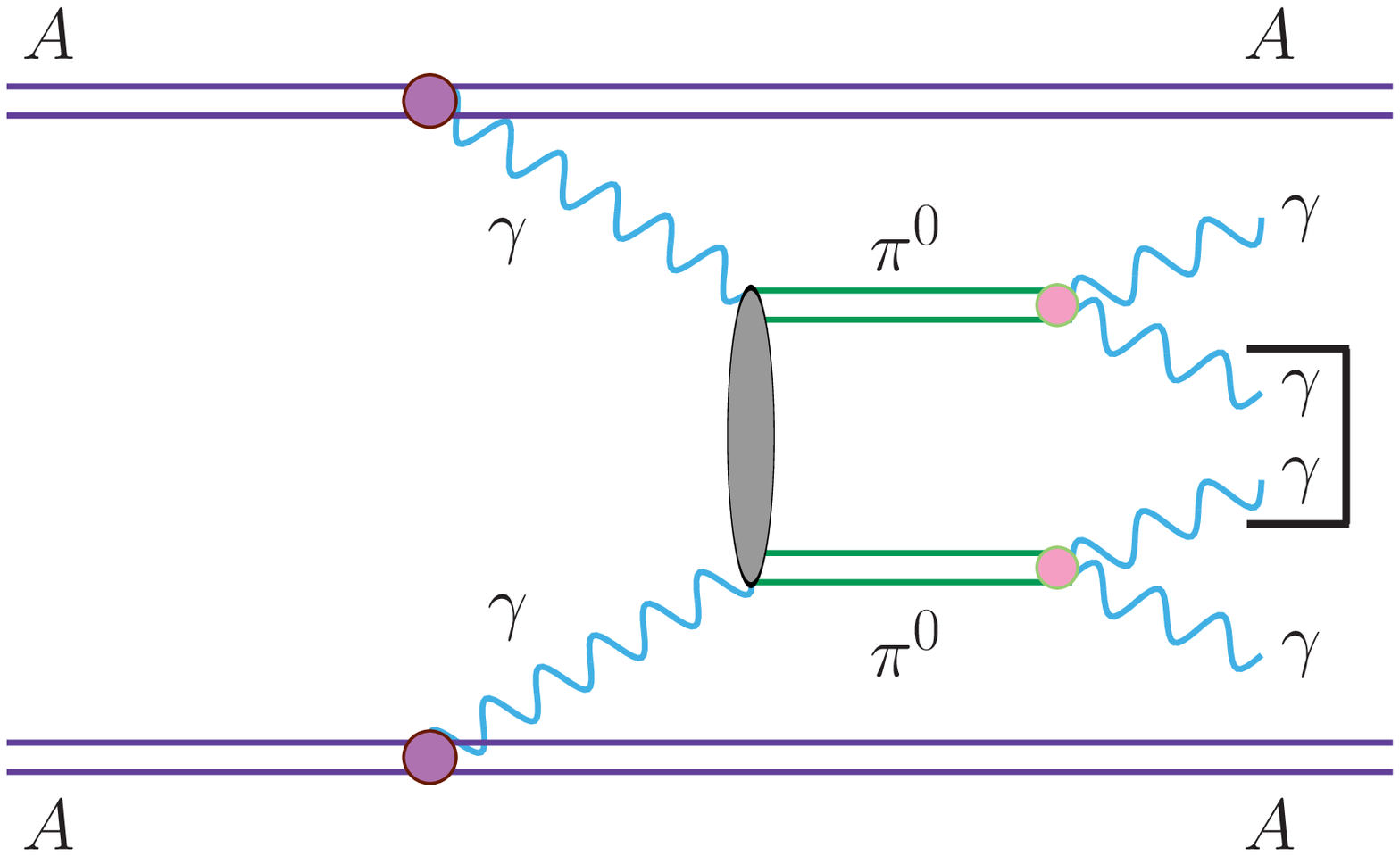}
	\caption{The continuum $\gamma \gamma \to \gamma \gamma$ scattering (a),
		$\gamma\gamma \to$ resonances $\to \gamma\gamma$ (b), 
		and the background mechanism (c).}
	\label{fig:diagrams}
\end{figure}

%----------------------------------------------------------------------------
We consider nuclear ultra-peripheral collisions (UPCs).
In Fig.~\ref{fig:diagrams} we illustrate the signal 
($\gamma \gamma \to \gamma \gamma$ scattering) which we take to be the dominant
box mechanism (see \cite{Klusek-Gawenda:2016euz}).
Panel (b) shows a diagram for $s$-channel $\gamma\gamma \to $
pseudoscalar/scalar/tensor resonances which also 
contributes to the
$\gamma \gamma \to \gamma \gamma$ process.
We also show (diagram (c)) the $\gamma\gamma \to \pi^0\pi^0$ process, which leads to what we consider as the dominant background
 when only one photon from each $\pi^0 \to \gamma\gamma$
decay is detected. 
Other processes such as diffractive multi-hadron production 
that result in only two measured photons may also
contribute to the background,
but can be reduced using the techniques shown here.
Being strongly dependent on the acceptance and detection thresholds,
these processes need to be experimentally assessed and are beyond
the scope of the current study.

%---------------------------------------------------
%---------------------------------------------------

In the equivalent photon approximation (EPA)
in impact parameter space,
the phase space integrated cross section for $A_1A_2 \to A_1 A_2 X_1 X_2$ reaction is expressed through the five-fold integral,
\begin{eqnarray}
\sigma_{A_1 A_2 \to A_1 A_2 X_1 X_2}\left(\sqrt{s_{A_1A_2}} \right) &=&
\int \sigma_{\gamma \gamma \to X_1 X_2} 
\left(W_{\gamma\gamma} \right)
N\left(\omega_1, {\bf b_1} \right)
N\left(\omega_2, {\bf b_2} \right)  \, S_{abs}^2\left({\bf b}\right) \nonumber  \\ 
& \times &
\mathrm{d}^2 b \, \mathrm{d}\overline{b}_x \, \mathrm{d}\overline{b}_y \, 
\frac{W_{\gamma\gamma}}{2}
\mathrm{d} W_{\gamma\gamma} \, \mathrm{d} Y_{X_1 X_2} \;,
\label{eq:EPA_sigma_final_5int}
\end{eqnarray} 
where $X_1X_2$ is a pair of photons or neutral pions.
$W_{\gamma\gamma}=\sqrt{4\omega_1\omega_2}$
and $Y_{X_1 X_2}=\left( y_{X_1} + y_{X_2} \right)/2$ 
are invariant mass and rapidity of the outgoing $X_1 X_2$ system. 
The energy of the photons is expressed through 
$\omega_{1/2} = W_{\gamma\gamma}/2 \exp(\pm Y_{X_1 X_2})$.
The variables ${\bf b_1}$ and ${\bf b_2}$ are impact parameters 
of the photon-photon collision point with respect to parent
nuclei 1 and 2, respectively, 
and ${\bf b} = {\bf b_1} - {\bf b_2}$ is the standard impact parameter 
for the $A_1 A_2$ collision.
The absorption factor $S_{abs}^2\left({\bf b}\right)$ assures UPC 
implying that the nuclei do not undergo nuclear breakup.
The photon fluxes ($N\left(\omega_i, {\bf b_i} \right)$) are expressed through a nuclear charge
form factor of the nucleus. In our calculations we use 
a realistic 
form factor which is a Fourier transform of the charge distribution 
in the nucleus.
More details can be found e.g. in Ref. \cite{KlusekGawenda:2010kx}.

%Eq.~(\ref{eq:EPA_sigma_final_5int}) 
%includes very important ingredient which determines a final state of considered process.
%This is an 

The elementary cross section $\sigma_{\gamma \gamma \to X_1 X_2}$ in 
Eq.~(\ref{eq:EPA_sigma_final_5int}) for
the $\gamma \gamma \to \gamma \gamma$ scattering is calculated 
within LO QED with fermion loops (see left panel of Fig.~1 in Ref.~\cite{Klusek-Gawenda:2016euz}).
The one-loop box diagrams were calculated with the help of 
the Mathematica package 
{\tt{FormCalc}} \cite{Hahn:1998yk} and the {\tt{LoopTools}}
library based on \cite{vanOldenborgh:1989wn} to evaluate one-loop integrals. 
In the numerical calculations we include box diagrams with leptons and quarks only.
Inclusion of the $W$-boson loops is not necessary because this contribution
is important only for energies larger than twice the $W$ boson mass \cite{Lebiedowicz:2013fta}.
%Used by us results were confronted with previous one \cite{Jikia:1993tc,Bern:2001dg,Bardin:2009gq}.
Our results have been compared to and agree with Ref. \cite{Jikia:1993tc,Bern:2001dg,Bardin:2009gq}.
Other production mechanisms were considered in Ref. \cite{Klusek-Gawenda:2016euz,Klusek-Gawenda:2016nuo}, but
their contributions are much smaller in the low diphoton 
mass region 
$M_{\gamma\gamma}<$ 5 GeV considered here.

	%-----------------------------------------------
	
	\subsection{Pion pair production} 
	\label{subsec:bkg}
	%-----------------------------------------------
	
	The elementary cross section for $\gamma \gamma \to \pi \pi$ was studied
	in detail in Ref.~\cite{Klusek-Gawenda:2013rtu}.
	Both $\gamma \gamma \to \pi^+\pi^-$ and $\gamma \gamma \to \pi^0\pi^0$
	reactions were considered within the physical framework describing existing
	experimental data. Two of us calculated, for the first time,  both the total
	cross section and angular distributions and demonstrated  significance of
	resonances, continuum and pQCD mechanisms in these processes. Following
	\cite{Klusek-Gawenda:2013rtu}, here we include nine resonances,
	$\gamma \gamma \to \pi^+\pi^- \to \rho^\pm \to \pi^0\pi^0$ continuum,
	Brodsky-Lepage and handbag mechanisms.  A detailed formalism and 
	description of these sub-processes can be found in Ref.
	\cite{Klusek-Gawenda:2013rtu}. The angular distribution for the
	$\gamma \gamma \to \pi^0\pi^0$ process can be written
	in standard form with the help of the $\lambda_1, \lambda_2$  photon helicity-dependent amplitudes,
	as a function of $z=\cos \theta$, where $\theta$ is the pion
	scattering angle
	\begin{equation}
	\frac{\mathrm{d} \sigma_{\gamma \gamma \to \pi^0 \pi^0} \left(W_{\gamma\gamma} \right)}{\mathrm{d}z} =
	\frac{\sqrt{\frac{W_{\gamma\gamma}^2}{4}-m_\pi^2}}{\frac{W_{\gamma\gamma}}{2}} \frac{4\pi}{4\times 64\pi^2W_{\gamma\gamma}^2}
	\sum_{\lambda_1,\lambda_2} \left|\mathcal{M}_{\gamma\gamma \to \pi^0\pi^0}\left(\lambda_1,\lambda_2 \right)\right|^2  \,.
	\end{equation}
	We use the formalism sketched above to calculate the multi-dimensional 
	distribution
	\begin{eqnarray}
	\frac{\mathrm{d} \sigma_{A_1 A_2 \to A_1 A_2 \pi^0_1 \pi^0_2}\left(\sqrt{s_{A_1A_2}} \right)}
	{\mathrm{d} y_{\pi^0_1} \, \mathrm{d} y_{\pi^0_2} \, \mathrm{d} p_{t,\pi^0}} &=&
	\int \frac{\mathrm{d} \sigma_{\gamma \gamma \to \pi^0_1 \pi^0_2} \left(W_{\gamma\gamma} \right)}{\mathrm{d}z} 
	N\left(\omega_1, {\bf b_1} \right)
	N\left(\omega_2, {\bf b_2} \right)  \, S_{abs}^2\left({\bf b}\right) \nonumber  \\ 
	& \times &
	\mathrm{d}^2 b \, \mathrm{d}\overline{b}_x \, \mathrm{d}\overline{b}_y \, 
	\frac{W_{\gamma\gamma}}{2}
	\frac{\mathrm{d} W_{\gamma\gamma} \, \mathrm{d} Y_{\pi^0_1 \pi^0_2}}{\mathrm{d} y_{\pi^0_1} \, \mathrm{d} y_{\pi^0_2} \, \mathrm{d} p_{t,\pi^0}} \, \mathrm{d}z \;,
	\label{eq:grid}
	\end{eqnarray}
	for the $A_1 A_2 \to A_1 A_2 \pi^0 \pi^0$ reaction.
	Here, $y_{\pi^0_1}$, $y_{\pi^0_2}$ are the rapidities of the first 
	and second pion,
	$p_{t,\pi^0}$ is the transverse momentum of the pions (identical in our LO approximation)
	and $z$ is the pion center-of-mass scattering angle.
	Integration of this formula allows comparison with Eq.~\eqref{eq:EPA_sigma_final_5int}.

	A dense three-dimensional grid of the triple
	differential cross section of Eq.~(\ref{eq:grid}) is prepared in a broad range of rapidities of both
	neutral pions and transverse momentum of one of them.
	A Monte Carlo code has been written to include radiative decays of both pions.
	As the pions are spin-0 mesons, the decays are taken to be isotropic in the rest frames
	of the decaying pions. Lorentz boosts are performed to obtain the kinematic distributions
	of photons in the laboratory (nucleus-nucleus center of mass) frame.
	Then logical conditions and cuts on the photons are imposed
	and distributions in different variables are obtained by an appropriate
	binning in the selected kinematic variable.
	The distributions are presented below and different experimental requirements are examined in order to estimate whether the $\gamma \gamma \to \gamma \gamma$ process can be observed.

%--------------------------------------------
%-----------------------------------------------

\subsection{Resonance contributions} 
\label{subsec:res}
%-----------------------------------------------

The angular distribution for the $s$-channel resonances 
is typically used in calculating Feynman 
diagram contributions in the form:
\begin{equation}
\frac{\mathrm{d} \sigma_{ \gamma\gamma \to R \to \gamma\gamma}(W_{\gamma\gamma})}{\mathrm{d} \cos \theta}  = \frac{1}{32\pi W_{\gamma\gamma}^2} \frac{1}{4} 
\sum_{\lambda_1, \lambda_2} 
\left| \mathcal{M}_{\gamma\gamma \to R \to \gamma\gamma}(\lambda_1, \lambda_2) \right|^2 \, ,
\label{eq:dsig_dz_res}
\end{equation}
where $\theta$ denotes the polar angle between the beam direction
and the outgoing nucleon in the c.m. frame,
$W_{\gamma\gamma}$ is the invariant mass of the $\gamma\gamma$ system.
The amplitudes for the $\gamma\gamma$ production through the $s$-channel
exchange of a pseudoscalar/scalar meson are written as
\begin{equation}
\mathcal{M}_{\gamma\gamma \to R \to \gamma\gamma}(\lambda_1, \lambda_2) = \frac{\sqrt{64\pi^2 W_{\gamma\gamma}^2 \Gamma^2_RBr^2(R \to \gamma\gamma)}}
{\hat{s}-m_R^2-im_R \Gamma_R} \times \frac{1}{\sqrt{2\pi}} \, \delta_{\lambda_1 - \lambda_2} \;.
\end{equation} 
Here we use the same notation as in Ref. \cite{Klusek-Gawenda:2013rtu}.
In the present analysis we take into account only pseudoscalar and scalar mesons:
$\eta$, $\eta'(958)$, $\eta_c(1S)$, $\eta_c(2S)$, $\chi_{c0}(1P)$.
Their masses $m_R$, total widths $\Gamma_R$ 
and branching ratios $Br(R \ \to \gamma\gamma)$ are taken from the PDG  
\cite{Patrignani:2016xqp}.

To calculate a resonance that decays into two photons, 
we use the following expression
\begin{eqnarray}
\frac{\mathrm{d} \sigma_{A_1 A_2 \to A_1 A_2 \gamma\gamma }\left(\sqrt{s_{A_1A_2}} \right)}
{\mathrm{d} y_{\gamma_1} \, \mathrm{d} y_{\gamma_2} \, \mathrm{d} p_{t,\gamma}} &=&
\int \frac{\mathrm{d} \sigma_{\gamma\gamma \to R \to \gamma\gamma} \left(W_{\gamma\gamma} \right)}{\mathrm{d} \cos \theta } 
N\left(\omega_1, {\bf b_1} \right)
N\left(\omega_2, {\bf b_2} \right)  \, S_{abs}^2\left({\bf b}\right) \nonumber  \\ 
& \times &
\mathrm{d}^2 b \, \mathrm{d}\overline{b}_x \, \mathrm{d}\overline{b}_y \, 
\frac{W_{\gamma\gamma}}{2}
\frac{\mathrm{d} W_{\gamma\gamma} \, \mathrm{d} Y_{\gamma_1 \gamma_2}}{\mathrm{d} y_{\gamma_1} \, \mathrm{d} y_{\gamma_2} \, \mathrm{d} p_{t,\gamma}} \, \mathrm{d} \cos \theta 
\label{eq:resonances_nuclear}
\end{eqnarray}
for the calculation of 
the nuclear cross section.
This formula includes the unpolarized differential 
elementary cross section of Eq.~\eqref{eq:dsig_dz_res}.  
Further discussions of the cross section 
for the mesonic resonant states in the context of
light-by-light scattering are in Ref. \cite{Lebiedowicz:2017cuq}. 
%In our study, we implement $\gamma \gamma \to R \to \gamma \gamma$
%in nuclear reactions.

%-----------------------------------
\section{Experimental acceptance and resolution}
\label{sec:acc}
%-----------------------------------

We briefly summarize the experimental acceptance for measuring two-photon
final states in Run 3 and beyond at the LHC, and describe the experimental
resolution used in deriving our results.
To illustrate our case, we take the acceptance of the ALICE central barrel
at mid-rapidity, and the LHCb spectrometer acceptance at forward rapidities.

The ALICE central barrel covers the pseudorapidity range $|\eta|<$ 0.9
\cite{ALICE:2014}. Within this range, ALICE is capable of measuring photons
by different methods. First, electromagnetic calorimeters
EMCal and PHOS cover part of the solid angle \cite{EMCal:2010,PHOS:2014}.
Second, photons can be detected by the photon conversion method
(PCM). In this approach, photons which convert into
$e^{+}e^{-}$ pairs in the detector material are reconstructed by detecting
the two charged tracks of the lepton pair. Whereas the photon measurement
can be carried out with high efficiency but limited solid angle by the
electromagnetic calorimeters, the measurement by PCM covers the full solid angle of the central
barrel but suffers from reduced efficiency.
Additionally, hybrid measurements are possible with
one photon being detected by the calorimeters, and the other photon
being reconstructed by the PCM.
Analyses of $\pi^{0}$ and $\eta$ meson production in
proton-proton collisions at $\sqrt s$ = 8 TeV by PCM
in the ALICE central barrel resulted in a mass resolution 
$\sigma_{M}$ of
the $\eta$ meson $\sigma_{M_{\eta}} \sim$~5~MeV
(see Fig.~4 in Ref.~\cite{Acharya:2017tlv}), from which a PCM energy 
resolution
for single photons of $\sigma_{E_\gamma}/E_\gamma \sim$~1.3~\%
is deduced. For the azimuthal angular resolution $\sigma_\phi$, we 
choose a value of $\sigma_\phi$ = 0.02 in order to illustrate the different 
behaviour of scalar and vector asymmetries explained and used below for suppressing
the background.
A comparative study of photon measurements by
    electromagnetic calorimeters to the PCM for measuring
    light-by-light scattering in the energy range considered,
  as well as a comprehensive analysis of the experimental
  resolutions, requires detailed studies of the response of the
  detectors used.  Such an analysis is beyond the scope of this paper.

LHCb is fully instrumented in the pseudorapidity range 
$2 < \eta < 4.5$ with tracking, calorimetry and particle identification.
Tracks can be reconstructed down to a transverse momentum of about 100~MeV,
and photons down to a transverse energy of about 200~MeV.   In this study
we assume that any photon with $E_{t,\gamma} > 200$ MeV  and $2 < \eta < 4.5$
will be detected by LHCb, while photons outside this region are undetected.
The energy resolution is parametrized as \cite{AbellanBeteta:2012dd}:
\begin{equation}
  \frac{\sigma_{E_\gamma}}{E_\gamma} =
  \frac{0.085}{\sqrt{E_\gamma}} +
  \frac{0.003}{E_\gamma}+ 0.008 \, ,
\label{LHCb_energy_resolution}
\end{equation}
where $E_\gamma$ is the photon energy in GeV.
These fiducial requirements and resolution allow us to make rough estimates for
the feasibility of observing light-by-light scattering:  a full study
using the LHCb detector simulation would be necessary to obtain precise results (Ref.~\cite{Clemencic:2011zza}).

The fiducial regions used in the present study are summarized in Table \ref{tab:cuts}.
In the following they will be named ``ALICE-fiducial'' or ``LHCb-fiducial''
for brevity. It should be noted that for massless
particles in the final state, $E_{t,\gamma}$ and $p_{t,\gamma}$, as well as rapidity and pseudorapidity are identical.
%---------------------------------------------------------------------------
\begin{table}[!h]
	\caption{Fiducial regions used in the present study.}
	\label{tab:cuts}
	\begin{tabular}{|c|c|c|}
		\hline
		experiment   &   pseudorapidity range  		&   energy cut    \\
		\hline
		ALICE        & -0.9 $< \eta_{\gamma} <$ 0.9	&   $E_{\gamma} >$ 0.2 GeV \\
		\hline
		LHCb         &  2.0 $< \eta_{\gamma} <$ 4.5	&   $E_{t,\gamma} >$ 0.2 GeV \\
	\hline
\end{tabular}
\end{table}
%--------------------------------------------------------------------------

%-----------------------------------
\section{Cross sections}
\label{sec:cross}
%-----------------------------------

We present the total cross section for different contributions  to the diphoton
final state in Table \ref{tab:events}.
In this table, the total cross section is listed for two ranges of photon-photon
invariant mass.  The first range is from 0 to 2 GeV, with the  second range 
for values $W_{\gamma\gamma}$ larger than 2 GeV.
We take $W_{\gamma\gamma}^{max}$= 50 GeV for
fermionic boxes,
$W_{\gamma\gamma}^{max}$ = 5 GeV for the $\pi^0\pi^0$ background (it is negligible above $W_{\gamma\gamma}$ = 5 GeV) and
$W_{\gamma\gamma} \in (m_R - 1 \mbox{ GeV}, m_R + 1 \mbox{ GeV})$ for resonances.

%---------------------------------------------------------------------------
\begin{table}[!h]
	\caption{Total nuclear cross section in nb 
		for the Pb-Pb collision energy 
			$\sqrt{s_{NN}} = 5.02$~TeV.}
	\label{tab:events}
	\begin{tabular}{|l|r|r|r|r|}
		\hline
		Energy   &  \multicolumn{2}{c|}{$W_{\gamma\gamma} = (0-2)$ GeV}  
		&  \multicolumn{2}{c|}{$W_{\gamma\gamma}>$ 2 GeV}   \\ \hline
		Fiducial region	& ALICE		& LHCb		& ALICE		& LHCb	\\ \hline \hline
		boxes       &   4 890 	& 3 818		& 146		& 79  	\\ \hline
		$\pi^0\pi^0$ background  & 135 300	& 40 866	& 46 		& 24	\\ \hline
		$\eta$		& 722 573	& 568 499	& 			&		\\ \hline
		$\eta'(958)$&  54 241	&  40 482	& 			&		\\ \hline
		$\eta_c(1S)$&			&			& 9			& 5		\\ \hline
		$\chi_{c0}(1P)$& 		&			& 4			& 2		\\ \hline
		$\eta_c(2S)$&			&			& 2			& 1		\\ \hline	 
	\end{tabular}
\end{table}
%--------------------------------------------------------------------------

The largest cross section is obtained for $\eta$ resonance production.
The background contribution dominates over the signal 
up to $W_{\gamma\gamma} = 2$~GeV, however, as discussed in the following, can be reduced.
%In view of larger masses of $\eta_c(1S)$, $\chi_{c0}(1P)$ and $\eta_c(2S)$ 
%resonance, the contribution from these resonant 
%mesonic states occurs only at second considered range of energy. 
Comparing results for $W_{\gamma\gamma}>2$~GeV, 
%	that are
%	obtained for ALICE and LHCb limitation,
the cross section for 
light-by-light scattering at midrapidity is
about a factor 2 larger than at forward rapidity.
%	In the case of lower invariant mass energies
%	(the first considered range of energy),
The cross sections for both fermionic box and resonant signal 
are similar
for the ALICE and LHCb fiducial regions. 
%It is essential to reduce the pionic background especially at midrapidity.

%--------------------
\subsection{Differential cross section at midrapidity}
\label{sec:diff-cross-midrap}
%--------------------

%-------------------------------------------------------------------------
%\begin{figure}[!h]
%	\includegraphics[width=0.45\linewidth]{dsig_dW_pi0pi0_ALICE_nm}
%	\caption{\label{fig:dsig_dM_ALICE} 
%		Invariant diphoton mass distribution for the standard ''ALICE cut''. 
%		The solid line is the signal due to Standard Model box contribution, 
%		while the dashed line corresponds to the $\pi^0 \pi^0$ background
%		defined in the main text.
%	}
%\end{figure} 
%-------------------------------------------------------------------------
%-----------------------------------------------
%------------------------------------------------

\begin{figure}[!h]
	\includegraphics[width=0.45\linewidth]{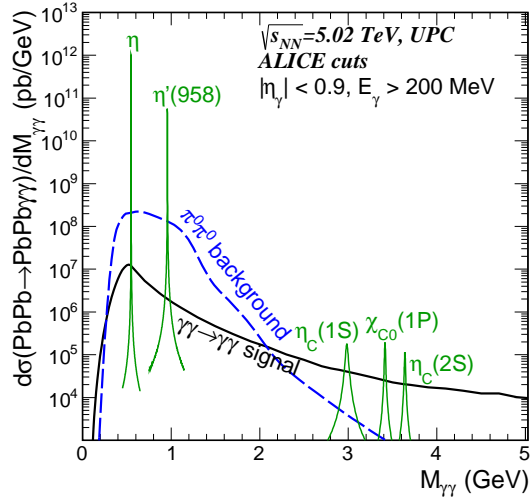}
	\caption{Differential cross section as a function of invariant diphoton mass within the ALICE fiducial region. }
	\label{fig:dsigma_dM_ALICE}
\end{figure}

First we present some distributions within the ALICE fiducial region.
The invariant mass distribution of two photon final states contains different
contributions of signal and background as discussed in
sections \ref{sec:theo}.
In Fig.~\ref{fig:dsigma_dM_ALICE}, the contribution to the signal  due to
the Standard Model boxes is shown by the solid black line. The contributions
to the signal by the different meson resonances are shown by the solid green
lines, while the dashed blue line represents the $\pi^0 \pi^0$ background as
discussed in section 2.1.
The $\pi^{0}\pi^{0}$ background shown in Fig.~\ref{fig:dsigma_dM_ALICE} is
composed of events where exactly two out of the four decay photons 
are within the fiducial volume, with the condition that each $\pi^{0}$ of 
the pair contributes one photon. At low invariant photon masses
$W_{\gamma\gamma} <$ 1.5 GeV, this background dominates over the signal
by about an order of magnitude but can be reduced by taking into account
the different phase space distribution of the signal and the background.

%Imposing some condition on the relative azimuthal angle
%or acoplanarity can help in eliminating the $\pi^0 \pi^0$ background 
%(see right panel).

%------------------------------------------------------------------------
%\begin{figure}[!h]
%	\includegraphics[width=0.45\linewidth]{dsig_dphid_ALICE_bg_nm}
%	\includegraphics[width=0.45\linewidth]{dsig_dW_pi0pi0_ALICE_phidcuts}
%	\caption{Background distribution in relative azimuthal angle between the two registered
%		photons for the ALICE kinematics.}
%	\label{fig:dsig_dphi_ALICE}
%\end{figure}
%-------------------------------------------------------------------------

%-------------------------------------------------------------------------
\begin{figure}[!h]
(a)	\includegraphics[width=0.45\linewidth]{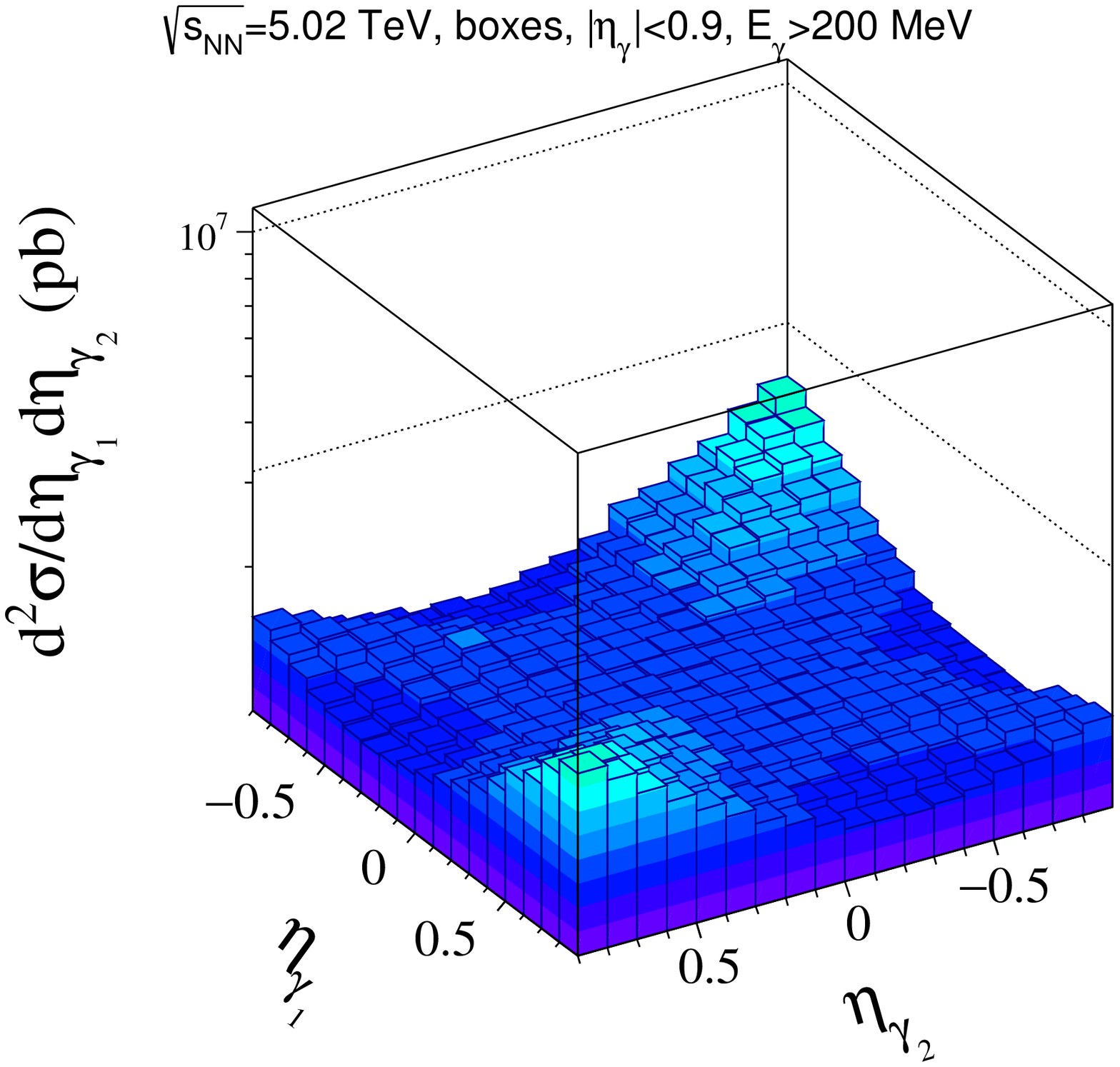}
(b)	\includegraphics[width=0.45\linewidth]{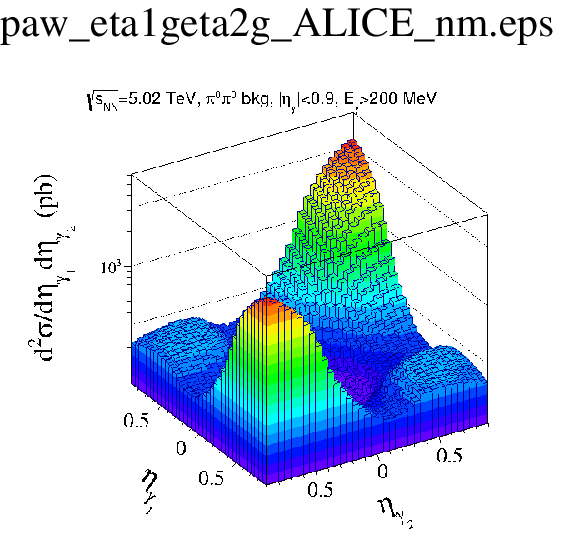}
	\caption{\label{fig:dsig_deta1deta2_ALICE} 
		Two-dimensional distribution of the signal (left panel) and $\pi^0\pi^0$ background
		(right panel) in rapidity of first and second photon (randomly chosen).
	}
\end{figure}
%-------------------------------------------------------------------------

%As an example of the different phase distribution of signal 
%(a) and background (b),
The two-dimensional distribution in rapidity of the first and second photon for signal and $\pi^0\pi^0$ background is shown in Fig.~\ref{fig:dsig_deta1deta2_ALICE} . 
The two distributions are 
peaked at forward and backward rapidities and are qualitatively rather similar, but differ by about two orders of magnitude.
This figure shows that cuts on
$\eta_{sum} = \eta_1+\eta_2$ could be used to reduce the background.
However this leads to marginal improvements due to the similarity
in the shape of signal and background. 
\subsection{Differential cross section at forward rapidity}
\label{sec:diff-cross-forward}
%-------------------

\begin{figure}[!h]
(a)	\includegraphics[width=0.45\linewidth]{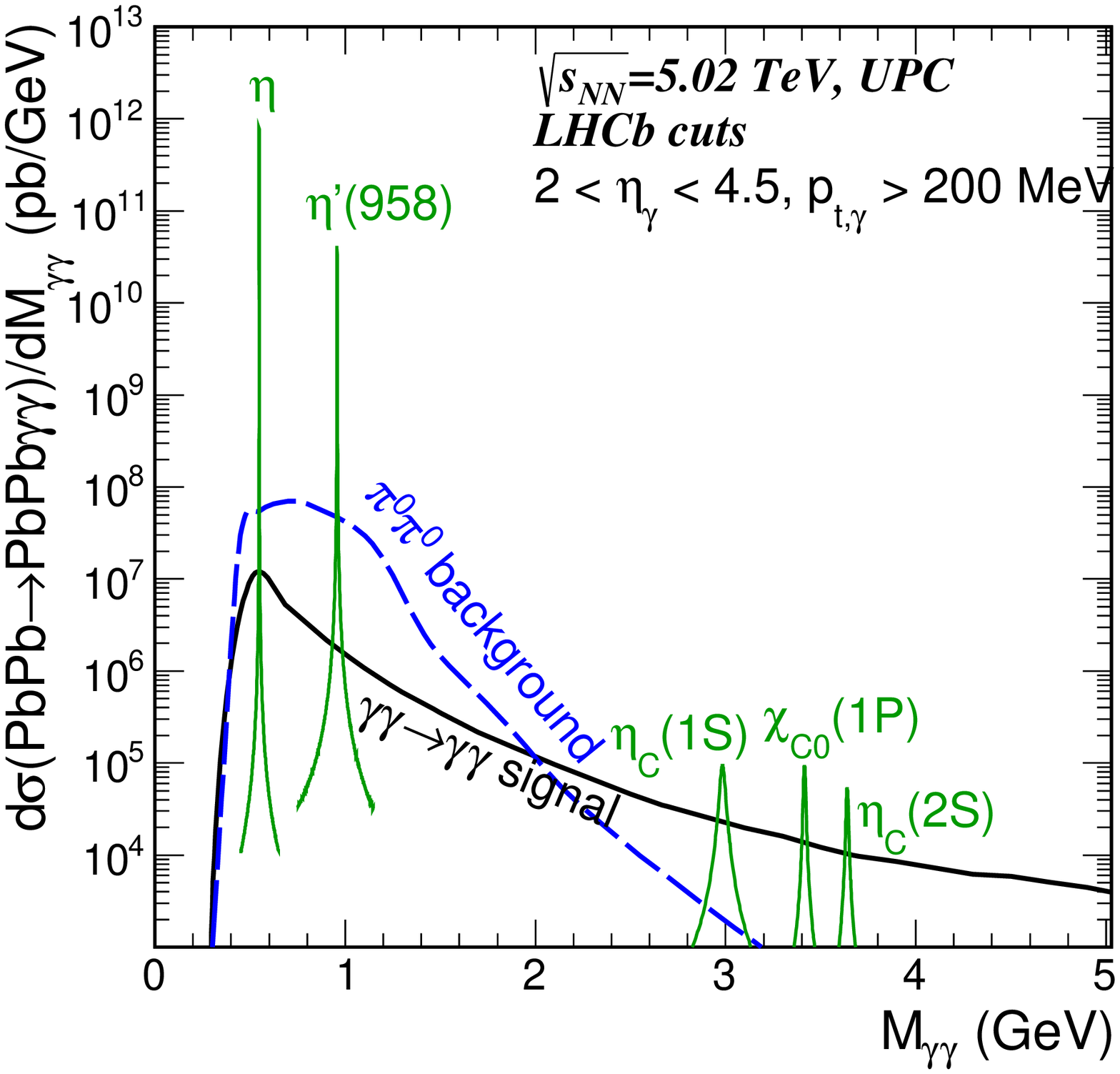}
(b)	\includegraphics[width=0.45\linewidth]{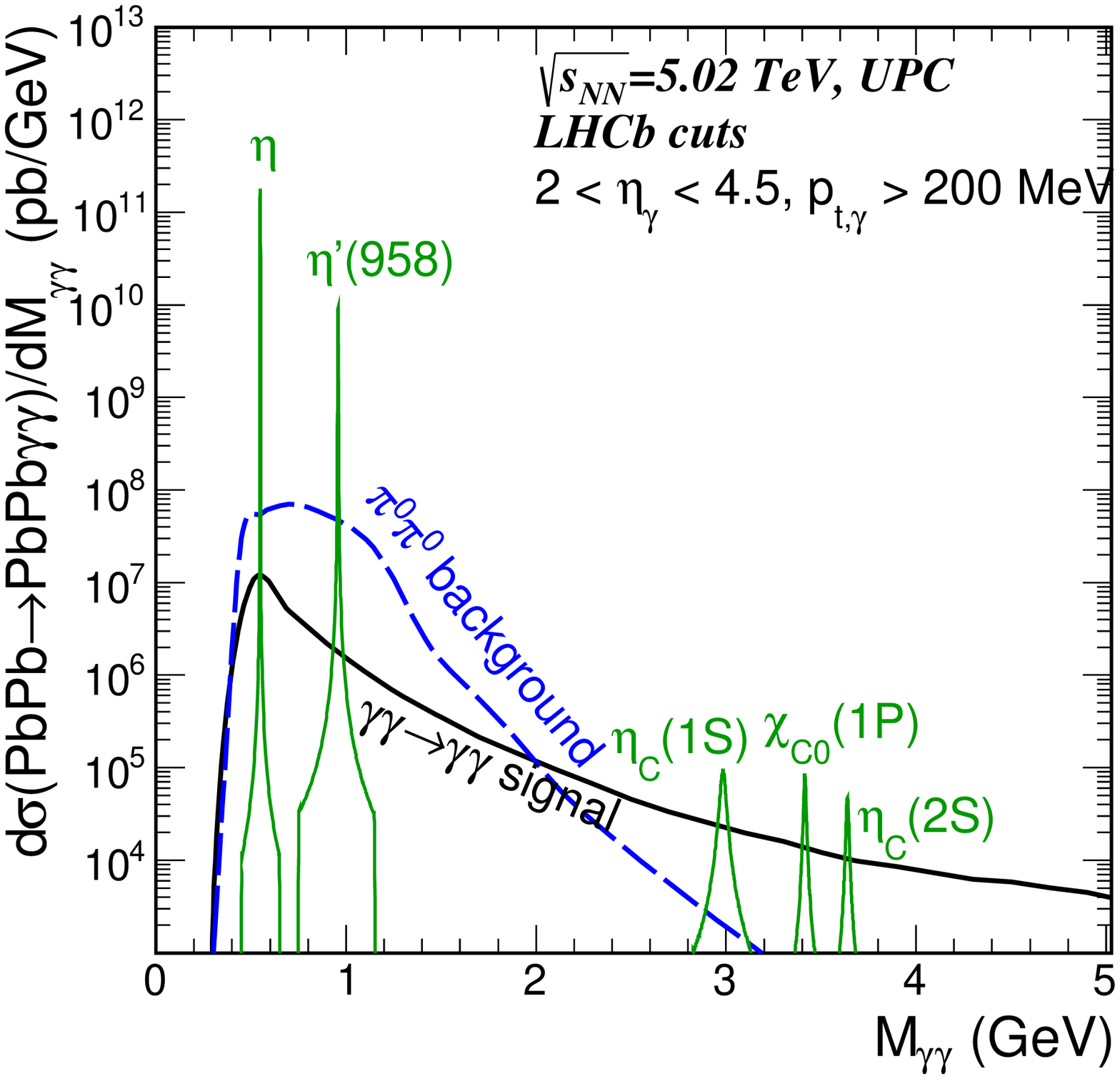}
	\caption{Invariant diphoton mass distribution for the standard LHCb 
		fiducial region presented without (a) and with (b)
			energy resolution.
	}
	\label{fig:dsigma_dM_LHCb}
\end{figure}

%The analysis of light-by-light scattering in ultra-peripheral heavy-ion collision at forward rapidity is realised using the LHCb acceptance. 
%For massless particles such as real photons, rapidity and pseudorapidity coincide.
Equivalent distributions are now shown within the LHCb fiducial region.
In Fig.~\ref{fig:dsigma_dM_LHCb} we show
the diphoton invariant mass distribution.
% within the LHCb fiducial region.
The distributions are similar to the ALICE conditions 
both in normalization and shape. 
The solid black line is the signal corresponding to the Standard Model box contribution, 
	the solid green lines correspond to
	resonant mesonic states
	while the dashed line corresponds to the $\pi^0 \pi^0$ background
	defined in the main text.
The relative background is slightly lower than for ALICE 
but the conclusion again is that 
the signal can be clearly observed only for $W_{\gamma\gamma} >$ 2 GeV
and one can observe very clear contributions coming from $\eta$ and $\eta'(958)$ resonances.
%It seems to be worth mentioning that peak heights of resonances very strongly depend on the number of bins. The maximum value of the differential cross section emerges exactly at $m_R$. 
The inclusion of the LHCb energy resolution (Eq.~\eqref{LHCb_energy_resolution}) broadens the peak 
in the distribution, which is plotted in Fig. 4 with and without energy smearing.  
%The peak height appears to decrease, although clearly the number of events in a window around the peak position remains the same.

%a rather significant impact on the resonant contributions. Then peak height of $\eta$ and $\eta'$ is almost one order of magnitude smaller. However, energy resolution does not modify the value of the total cross section.    

%-------------------------------------------------------------------------
%\begin{figure}[!h]
%	\includegraphics[width=0.45\linewidth]{dsig_dW_pi0pi0_LHCb_nm}
%	\caption{\label{fig:dsig_dM_LHCb} 
%		Invariant diphoton mass distribution for the LHCb cut. The solid line
%		is the signal due to the Standard Model box contribution, while the dashed
%		line corresponds to the $\pi^0 \pi^0$ background.
%	}
%\end{figure}
%-------------------------------------------------------------------------

%-----------------------------------------------
%------------------------------------------------

%------------------------------------------------------------------------
%\begin{figure}[!h]
%	\includegraphics[width=0.45\linewidth]{dsig_dphid_LHCb_bg_nm}
%	\caption{\label{fig:dsig_dphi_LHCb}
%		Distribution in relative azimuthal angle between the two registered
%		photons for the LHCb kinematics.
%	}
%\end{figure}
%--------------------------------------------------------------------------

%Azimuthal angle correlations are shown in Fig.\ref{fig:dsig_dphi_LHCb}.
%Again similar conclusions as for the ALICE setup can be drawn.
%The photons of the background are not back-to-back. 
%Here we do not show the result for fermionic boxes because,
%on the theoretical side, this distribution gives a peak exactly at 180$^o$.
%In Sec. \ref{sub:exp_res} we will include experimental resolution effects,
%but here we assume that photons from the $\gamma\gamma \to \gamma\gamma$ signal
%are very strongly correlated back-to-back. 

\begin{figure}[!h]
(a)	\includegraphics[width=0.45\linewidth]{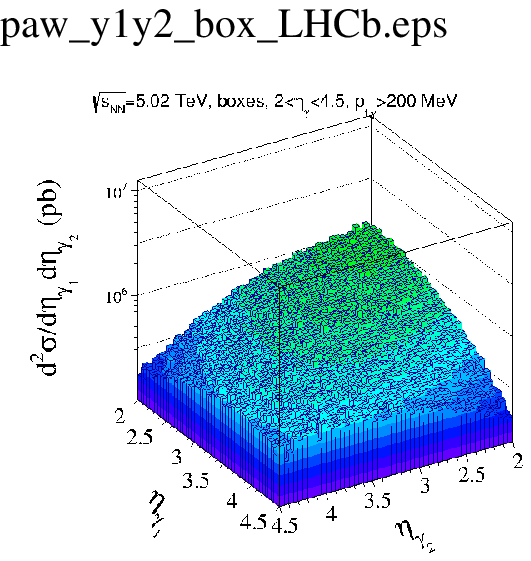}
(b)	\includegraphics[width=0.45\linewidth]{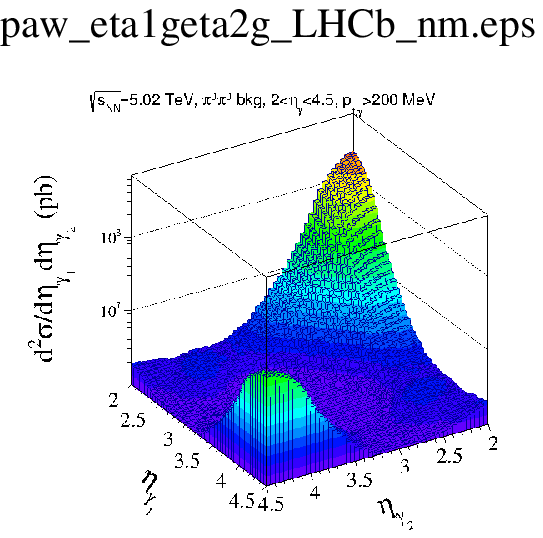}
	\caption{\label{fig:dsig_deta1deta2_LHCb} 
		Two-dimensional distribution of the signal (left panel) and background
		(right panel) in rapidity of first and second registered photon (chosen randomly).
	}
\end{figure}

In Fig.~\ref{fig:dsig_deta1deta2_LHCb} we show two-dimensional distributions
as a function of pseudorapidity of the first and second photon.
The left panel corresponds to the $\gamma\gamma \to \gamma\gamma$ (box) signal
and the right panel shows the result for the $\pi^0\pi^0$ background.
In some regions of ($\eta_1 \times \eta_2$) space the background contribution
is much larger than the signal one.
Note the different scale on $\mathrm{d}^2 \sigma /
\mathrm{d} \eta_{\gamma_1} \mathrm{d} \eta_{\gamma_2} $-axis
in the left and right panels. As in the case of
the ALICE fiducial region (Fig.~\ref{fig:dsig_deta1deta2_ALICE}),
the signal is two orders of magnitude smaller than the background
but in the LHCb case the shapes of the distributions are rather different.
%Here, in contrast to distribution for background, the largest cross section for signal occurs on the diagonal, not only in the case when two photons very close to limiting values of pseudorapidity are detected.  
Here there is relatively more background when
both photons have large rapidities.

%-------------------------------------------------------------------------

\section{Background suppression of $\pi^{0}\pi^{0}$ decays}

%\subsection{Background suppression}

The background from the $\pi^{0}\pi^{0}$ decays shown in
Fig.~\ref{fig:dsigma_dM_ALICE} can be reduced by kinematic cuts.
%We have checked validity of employing the $\eta_1+\eta_2$ limitation. However, this did not give a satisfactory level of reduction of	the background. 
%In Ref.~\cite{Klusek-Gawenda:2018ijg} it was shown that one can observe very limited region of very small $p_{t,\gamma \gamma} = |\overrightarrow{p_{t1}} + \overrightarrow{p_{t2}}|$
%where the $\gamma\gamma \to \gamma \gamma$ signal is above the pionic background.
%\st{while for the background, although the $\pi^0\pi^0$ system also has 
%very low $p_{t,\gamma}$, two of the four photons, in general, will not}
%\cite{Klusek-Gawenda:2018ijg}.
Not taking into account the experimental resolution, the two final state
photons of $\gamma\gamma\rightarrow\gamma\gamma$ scattering, shown in
Fig.~\ref{fig:diagrams} a,b), are of equal transverse momenta and are
back-to-back in azimuthal angle. 
In first order, 
the correlation in transverse momentum will be smeared out by the
experimental resolution in the measurement of the two photons. 
Higher order corrections, such as the finite values of the beam 
emittance and the crossing angle of
the colliding beams, are beyond the scope of the study presented here, and are
hence neglected in the results presented below.
The correlations of the signal can be quantified by two asymmetries,

\begin{equation}
A_{S}=\left|\frac{|\vec{p}_T(1)|-|\vec{p}_T(2)|}{|\vec{p}_T(1)|+|\vec{p}_T(2)|}\right| \, ,
\label{asyms}  
\end{equation}  
\begin{equation}
A_{V}=\frac{|\vec{p}_T(1)-\vec{p}_T(2)|}{|\vec{p}_T(1)+\vec{p}_T(2)|}
\label{asymv}  \, .
\end{equation}  

Here, the scalar asymmetry $A_{S}$ is a measure of the relative difference in
transverse momentum of the two photons, and is non-zero for two back-to-back
photons of the signal due to the finite energy resolution of the measurement.
The vector asymmetry $A_{V}$ defined in \mbox{Eq.~\eqref{asymv}} reflects a
convolution of the experimental resolutions of photon energy and azimuthal
angle measurement. The two-photon background resulting from
the $\pi^{0}\pi^{0}$ decays does not show the correlations discussed above.
The correlations can hence be used to suppress
the background by kinematical cuts on these two asymmetries.

\begin{figure}[!h]
	\centering
%(a)	\includegraphics[width=0.45\linewidth]{paw_asymm1asymm2_sig_ALICE_positive}
%(b)	\includegraphics[width=0.45\linewidth]{paw_asymm1asymm2_bkg_ALICE_positive}
(a)	\includegraphics[width=0.45\linewidth]{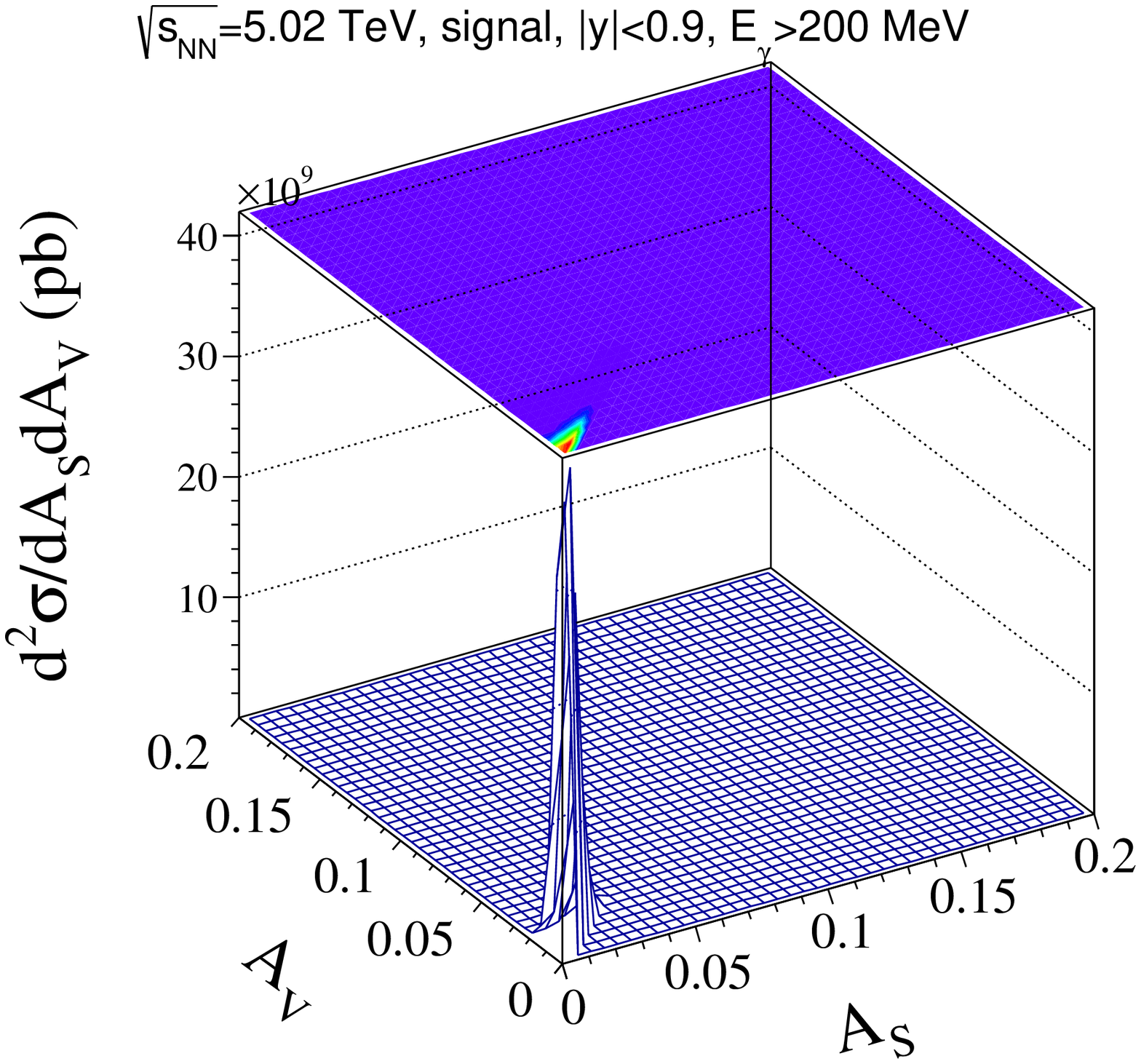}
(b)	\includegraphics[width=0.45\linewidth]{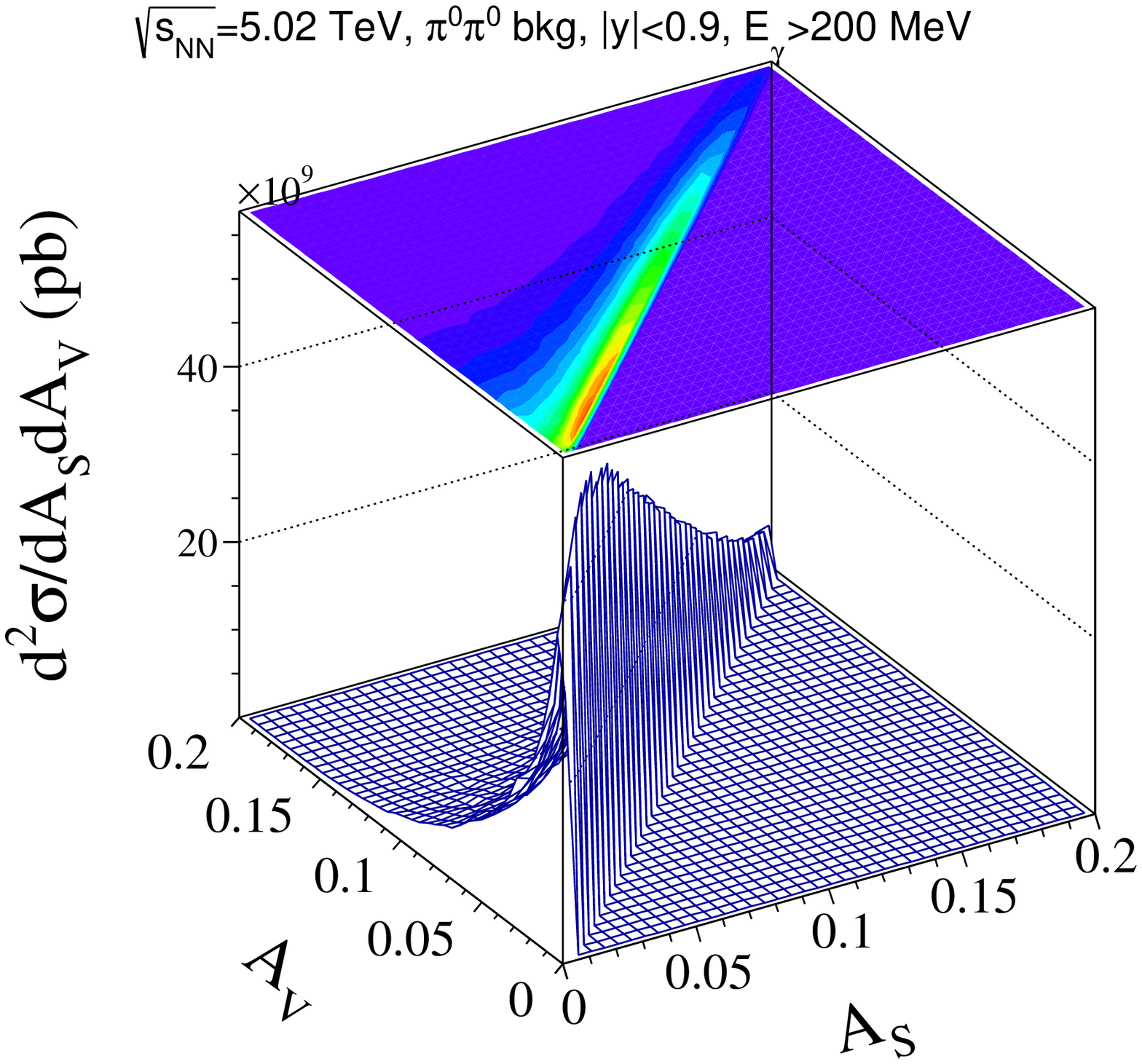}
	\caption{Vector vs scalar asymmetry of signal (left panel) and background (right panel).}
	\label{fig:asym2d}
\end{figure}

The correlation of the two asymmetries discussed above are shown for the
signal in Fig.~\ref{fig:asym2d}~(a) for an energy resolution of
$\sigma_{E_\gamma}/E_\gamma$ = 1.3~\% and an azimuthal angle
resolution of $\sigma_{\phi}$ = 0.02. From the considerations on asymmetries
outlined above, one expects the vector asymmetry to be larger than the scalar
asymmetry, $A_{V} > A_{S}$, as shown in Fig.~\ref{fig:asym2d}~(a). 
The corresponding histogram (Fig.~\ref{fig:asym2d}~(b))
is shown for photon background pairs resulting
from the $\pi^{0}\pi^{0}$ decays. This background distribution is an order of magnitude
wider compared to the distribution of the signal. A carefully chosen cut on the
asymmetry parameters $A_{S}$ and $A_{V}$ will therefore reduce the background
substantially, while reducing the signal only marginally.

\begin{figure}[!h]
	\centering
	\includegraphics[width=0.45\linewidth]{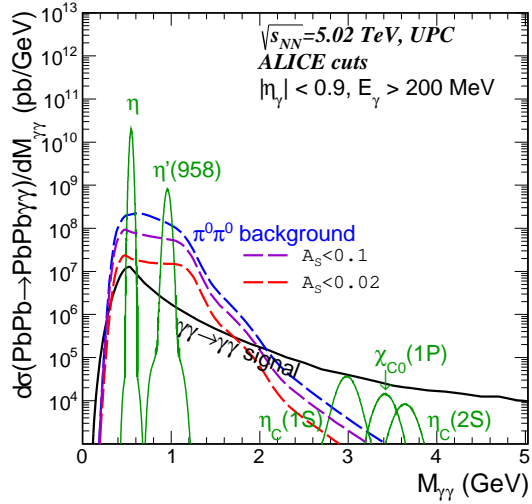}
	\caption{Signal and background with asymmetry conditions.}
	\label{fig:sigback_asymcut}
\end{figure}

The diphoton mass distribution is shown in Fig.~\ref{fig:sigback_asymcut}
from a study performed within the ALICE fiducial region,
with successive cuts on $A_S$ applied. The signal is reduced by a negligible amount for both values of $A_S$. 
The cut $A_S < 0.02$ reduces the background by about 
a factor of 10 as shown by the red line, and results in a remaining background  which is  a 
factor of about 10 larger than the signal at diphoton masses $M_{\gamma\gamma} \sim 1.2$ GeV.

	\begin{figure}[!h]
		\centering
		(a)	\includegraphics[width=0.45\linewidth]{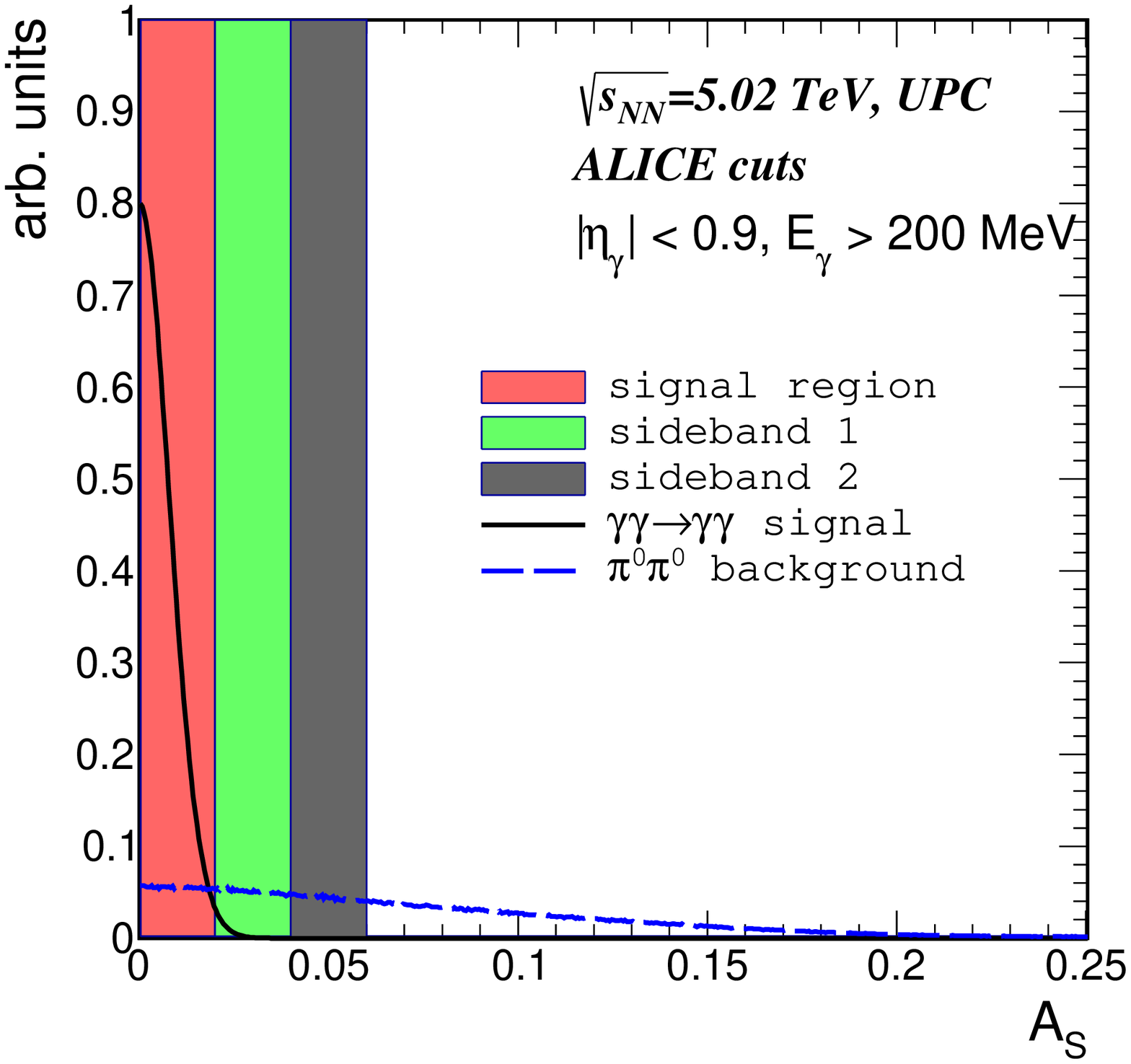}
		(b)	\includegraphics[width=0.45\linewidth]{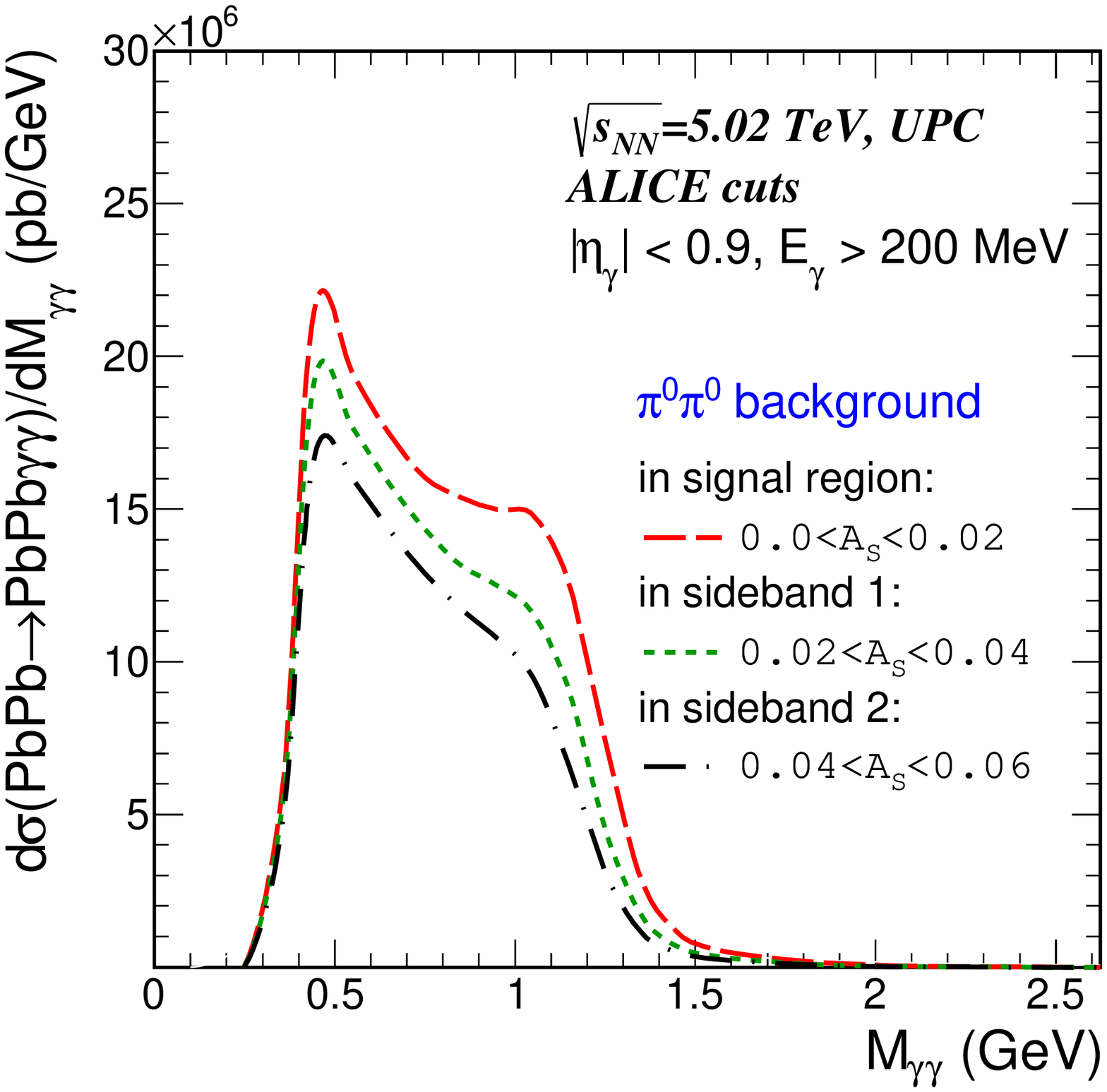}
		\caption{Scalar asymmetry distribution of signal and background
			(a), and signal and sideband background mass distribution (b).}
		\label{fig:sideband}
	\end{figure}

	A sideband subtraction method using the asymmetry parameter $A_{S}$ as
	separation variable can be used to extract the signal in this mass range.
	In this approach, kinematic distributions of signal and background, which can be 
		one or multi-dimensional,
	are fit in intervals of the separation variable. 
	For the diphoton signal a one-dimensional
	analysis could be based on the mass distribution 
	while a higher-dimensional
	approach could include the pseudorapidity distributions of
	the two photons shown in Fig. 3, and the transverse momentum distribution
	of the photon pair. To illustrate the one-dimensional approach,
	we show in Fig. 8~(a) the distribution of signal and background for the first
	three intervals of the separation variable, labeled signal region, sideband 1 and
	sideband 2. The signal region, defined by $0 < A_{S} < 0.02$, contains
	$\sim$ 95~\% of the signal events.  Sideband 1 and sideband 2 are defined
	by $0.02 < A_{S} < 0.04$  and $0.04 < A_{S} < 0.06$, respectively.  The
	distribution of the signal is at the 5~\% level in sideband 1, and negligibly
	small in sideband 2,  whereas the background distribution extends well into
	the sideband regions as shown in  Fig.~\ref{fig:sideband}~(a) by the
	dashed blue line. In  Fig.~\ref{fig:sideband}~(b), the mass distributions of
	background events in the signal  and the two sideband regions are shown by
	the red-dashed,  green-dotted and black-dot-dashed lines, respectively.  Clearly visible in this panel
	is a continuous reduction of the background cross section from the signal to
	the sideband 2 region, as expected by the behaviour of the background scalar
	asymmetry shown  in Fig.~\ref{fig:sideband}~(a).
%	This sideband subtraction technique allows to derive a quality
%	factor whether a measured diphoton event originates from the signal sample \textcolor{red}{\cite{Williams:2008sh}}.
Based on the kinematical fits of the signal and background distributions, the sideband analysis derives a factor expressing the likelihood of the event belonging to the signal or background sample (see e.g. \cite{Williams:2008sh}).
	Quantitative results on the misidentification of background as signal events
	depend, however, on the event statistics available 
	and information
	on the single photon detection efficiency, and are
	beyond the scope of the study presented here.

%-------------------------
\section{Conclusions}
%-------------------------

Ultra-peripheral heavy-ion collisions at high energies open the possibility to measure $\gamma \gamma \to \gamma \gamma$ scattering. 
So far the ATLAS and CMS collaborations obtained 
first evidence of photon-photon scattering for
invariant masses $W_{\gamma\gamma} >$ 6 and 5 GeV, respectively.
Due to the experimental cuts on transverse photon momenta 
$p_{t,\gamma} >$ 3 GeV, the resulting statistics is so far rather limited. 
The ATLAS result is roughly consistent with 
the Standard Model predictions for elementary cross section
embedded into state-of-art nuclear calculation 
including realistic photon fluxes as the
Fourier transform of realistic charge distribution.

Here we consider the possibility to study elastic $\gamma \gamma \to$
$\gamma \gamma$ scattering in the diphoton mass range
$W_{\gamma\gamma} < 5$ GeV at the LHC using ALICE and LHCb detectors.
Our results show that the contributions of the pseudoscalar resonances
$\eta$, $\eta^{'}(958)$ are clearly visible on top of the diphoton
mass continuum arising from lepton loop diagrams.
We have made first predictions for cross sections 
as a function of diphoton mass 
for typical acceptances of the ALICE and LHCb experiments.
The evaluation of counting rates needs, however, Monte Carlo simulations
which take into account detailed acceptances and realistic responses of the
detectors used for measuring two-photon final states.

In addition to the signal ($Pb Pb \to Pb Pb \gamma \gamma$)
we consider the  background dominated by the $Pb Pb \to Pb Pb \pi^0 \pi^0$
reaction when only two out of the four decay photons in the final state 
are registered. This background can be reduced by imposing cuts on scalar
and vector asymmetry of transverse momenta of the two photons.
Cuts on sum of photon rapidities (or the rapidity of the diphoton system)
can additionally be used to reduce the background.
The background remaining after these cuts dominates the 
signal by
a factor of about ten at diphoton masses $M_{\gamma\gamma} \sim$ 1.2 GeV.
The extraction of the signal in this mass range is feasible
by a multi-variate sideband subtraction analysis. Quantitative
results on the signal efficiency and background suppression
in this sideband subtraction approach depend on the statistics of
the data sample available for analysis.

In our study we take the dominant background shown in Fig.~\ref{fig:diagrams}~(c)
as arising from the decay of two $\pi^{0}$'s which are correlated by the
emission from two vertices which, however, cannot be resolved at the macroscopic
level due to detector resolution limitation. In a crossing of heavy-ion bunches,
single $\pi^{0}$ production can occur by different pairs of particles in the two
beams, resulting in multiple uncorrelated $\pi^{0}$ production. These $\pi^{0}$'s
emerge from different vertices which are spread along the interaction region of
the two colliding beams. The contribution of the decay photons of these
uncorrelated $\pi^{0}$'s to the background depends on the beam parameters at the
interaction point and the position resolution of the photon reconstruction
along the beam direction, and is beyond the scope of the study presented here.

%-------------------------
%\section{Acknowledgements}
%-------------------------
\acknowledgments

The authors thank the ExtreMe Matter Institute EMMI for the
support of the workshop "Challenges in Photon Induced Interactions"
in Krakow where this study was initiated.
This work has been supported by the Polish National Science Center 
grant DEC-2014/15/B/ST2/02528 (MKG and AS)
and by the German Federal Ministry
	of Education and Research under promotional reference 05P19VHCA1 (RS). 

%------------------------------------------------------------------
\bibliography{refs}
%------------------------------------------------------------------
\end{document}